\newif\iffinal
\finaltrue
\documentclass[nonacm,sigplan]{acmart}

\usepackage{tikz}
\usepackage{pifont}
\newcommand{\xmark}{\ding{55}}%
\usepackage{filecontents}

\usepackage{caption}
\usepackage{subfigure}
\usepackage{booktabs} 
\usepackage{enumitem}
\usepackage{balance}
\usepackage{graphicx}
\usepackage{amsmath,lipsum, amsthm}
\usepackage{fixltx2e}
\usepackage{url}
\usepackage{color}
\usepackage{listings}
\usepackage{xspace}
\usepackage{makecell}
\usepackage{fancyvrb}
\usepackage{verbatim}
\usepackage{etoolbox}
\usepackage{multirow}
\usepackage{comment}
\usepackage{hyperref}
\usepackage[normalem]{ulem}
\usepackage[noend]{algpseudocode}
\usepackage{algorithm}
\usepackage[normalem]{ulem}
\usepackage{float}
\usepackage{outlines}
\usepackage{cleveref}

\setlist{nosep}
\newcommand{\gpp}[0]{\textproc{GPP}\xspace}
\newcommand{\spp}[0]{\textproc{SPP}\xspace}
\newcommand{\ours}[0]{\textproc{GraphPipe}\xspace}
\newcommand{\pd}[0]{PipeDream\xspace}
\newcommand{\m}[1]{\mathcal{#1}}

\theoremstyle{plain}

\theoremstyle{definition}

\theoremstyle{remark}

\begin{document}
\title{\ours: Improving Performance and Scalability of DNN Training with Graph Pipeline Parallelism}


\author{Byungsoo Jeon}
\authornotemark[1]
\email{byungsoj.com@gmail.com}
\affiliation{%
  \institution{NVIDIA}
  \city{Arlington}
  \state{Virginia}
  \country{USA}
}

\author{Mengdi Wu}
\authornotemark[1]
\email{mengdiwu@andrew.cmu.edu}
\affiliation{%
  \institution{Carnegie Mellon Univerisity}
  \city{Pittsburgh}
  \state{PA}
  \country{USA}
}

\author{Shiyi Cao}
\authornotemark[1]
\email{shicao@berkeley.edu}
\affiliation{%
  \institution{UC Berkeley}
  \city{Berkeley}
  \state{CA}
  \country{USA}
}

\author{Sunghyun Kim}
\authornote{Equal contribution.} 
\email{sunghyun@csail.mit.edu}
\affiliation{%
  \institution{MIT}
  \city{Cambridge}
  \state{MA}
  \country{USA}
}

\author{Sunghyun Park}
\email{sunghyunp@nvidia.com}
\affiliation{%
  \institution{NVIDIA}
  \city{Seattle}
  \state{Washington}
  \country{USA}
}

\author{Neeraj Aggarwal}
\email{aggarwal.neeraj141@gmail.com}
\affiliation{%
  \institution{Carnegie Mellon Univerisity}
  \city{Pittsburgh}
  \state{PA}
  \country{USA}
}

\author{Colin Unger}
\email{unger@stanford.edu}
\affiliation{%
  \institution{Stanford University}
  \city{Palo Alto}
  \state{CA}
  \country{USA}
}

\author{Daiyaan Arfeen}
\email{marfeen@andrew.cmu.edu}
\affiliation{%
  \institution{Carnegie Mellon Univerisity}
  \city{Pittsburgh}
  \state{PA}
  \country{USA}
}

\author{Peiyuan Liao}
\email{peiyuanl@andrew.cmu.edu}
\affiliation{%
  \institution{Carnegie Mellon Univerisity}
  \city{Pittsburgh}
  \state{PA}
  \country{USA}
}

\author{Xupeng Miao}
\email{xupeng@cmu.edu}
\affiliation{%
  \institution{Carnegie Mellon Univerisity}
  \city{Pittsburgh}
  \state{PA}
  \country{USA}
}

\author{Mohammad Alizadeh}
\email{alizadeh@csail.mit.edu}
\affiliation{%
  \institution{MIT}
  \city{Cambridge}
  \state{MA}
  \country{USA}
}

\author{Gregory R. Ganger}
\email{ganger@andrew.cmu.edu}
\affiliation{%
  \institution{Carnegie Mellon Univerisity}
  \city{Pittsburgh}
  \state{PA}
  \country{USA}
}

\author{Tianqi Chen}
\email{tqchen@cmu.edu}
\affiliation{%
  \institution{Carnegie Mellon Univerisity}
  \city{Pittsburgh}
  \state{PA}
  \country{USA}
}

\author{Zhihao Jia}
\email{zhihao@cmu.edu}
\affiliation{%
  \institution{Carnegie Mellon Univerisity}
  \city{Pittsburgh}
  \state{PA}
  \country{USA}
}


\begin{abstract}

Deep neural networks (DNNs) continue to grow rapidly in size, making them infeasible to train on a single device.
Pipeline parallelism is commonly used in existing DNN systems to support large-scale DNN training by partitioning a DNN into multiple stages, which concurrently perform DNN training for different micro-batches in a pipeline fashion. However, existing pipeline-parallel approaches only consider {\em sequential} pipeline stages and thus ignore the topology of a DNN, resulting in missed model-parallel opportunities.

This paper presents {\em graph pipeline parallelism} (\gpp), a new pipeline-parallel scheme that partitions a DNN into pipeline stages whose dependencies are identified by a directed acyclic graph. \gpp generalizes existing sequential pipeline parallelism and preserves the inherent topology of a DNN to enable concurrent execution of computationally-independent operators, resulting in reduced memory requirement and improved GPU performance.
In addition, we develop \ours, a distributed system that exploits \gpp strategies to enable performant and scalable DNN training. \ours partitions a DNN into a graph of stages, optimizes micro-batch schedules for these stages, and parallelizes DNN training using the discovered \gpp strategies. Evaluation on a variety of DNNs shows that \ours outperforms existing pipeline-parallel systems such as PipeDream and Piper by up to 1.6$\times$. \ours also reduces the search time by 9-21$\times$ compared to PipeDream and Piper.


\end{abstract}

\maketitle
\pagestyle{plain}


\section{Introduction} \label{sec:intro}

Deep neural networks (DNNs) grow more rapidly in size against hardware developments, making them computationally costly to train~\cite{openai2018computecost, patterson2021carbon}. A recent language model GPT-4 \cite{openai2023gpt4} supposedly uses a much larger number of parameters \cite{mittechreview2023gpt4size} compared to the previous model GPT-3 with 175 billion parameters~\cite{brown2020language}. 
As a result, training modern DNNs requires distributing the model architecture across multiple devices.

\begin{figure}[t!]
    \centering
    \includegraphics[width=0.8\columnwidth]{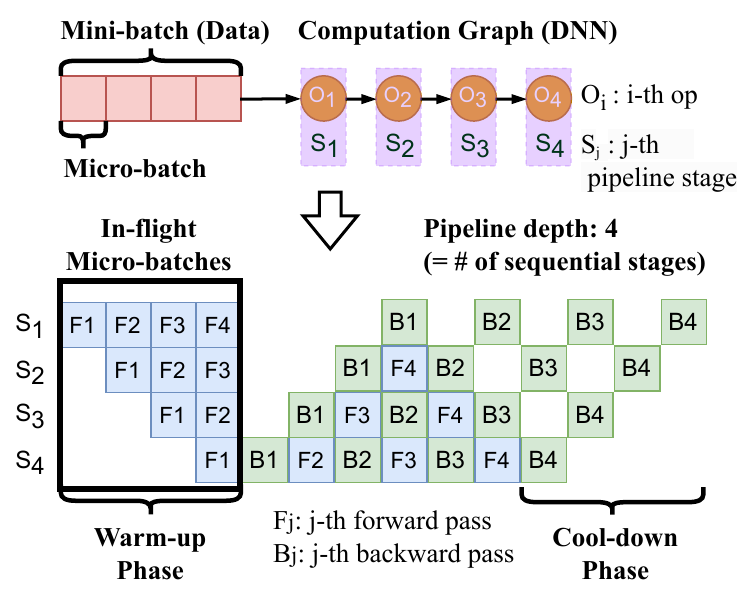}
    \caption{Pipeline parallelism for DNN training with basic terms used in this paper.}
    \label{fig:pp}
\end{figure}

To address this challenge, existing DNN systems apply model parallelism \cite{dean2012large, shazeer2018mesh, shoeybi2019megatron, lepikhin2020gshard, narayanan2021efficient, zheng2022alpa} where a DNN is partitioned into smaller pieces, each of which fits into the memory of a single device.
Pipeline parallelism \cite{huang2019gpipe, fan2021dapple, narayanan2021memory, narayanan2019pipedream, tarnawski2021piper} is a particular form of model parallelism that further improves device utilization and throughput. As shown in \Cref{fig:pp}, a key idea of pipeline parallelism is to split both a DNN and a mini-batch of samples into smaller pieces. First, the DNN is partitioned into multiple disjoint {\em stages}, each of which is a sub-model and links to other stages to form a pipeline. Second, a mini-batch of samples is further divided into multiple {\em micro-batches}, which are executed on different stages in a pipeline fashion. This 
approach reduces device idle time in training iterations, during each of which a single data mini-batch is processed, and thus improves throughput.

\paragraph{Shortcomings of existing sequential pipeline parallelism.} Existing schemes of applying pipeline parallelism form a \emph{sequential} pipeline from partitioned stages, which we refer to as \emph{sequential pipeline parallelism} (\spp).

\Cref{fig:pp} illustrates a DNN training scheme that employs it. A micro-batch traverses the pipeline's stages ($S_1$ to $S_4$) in sequence to perform the computations ($O_1$ to $O_4$) dictated by the DNN (forward pass: $F_1$'s), and traverses in reverse for all stages to update their assigned model weights (backward pass: $B_1$'s). Each stage needs to store the intermediate activations of a forward pass until its corresponding backward pass is completed. For a given stage, a micro-batch is \emph{in-flight} until its backward pass finishes. As micro-batches are continuously injected into the pipeline, there is a warm-up of in-flight micro-batches. The earlier the stage in the pipeline, the longer the warm-up. As described, \spp is simple to construct and operate, but has three key limitations.

First, opportunities to exploit the inherent parallel structures of a DNN are left unseized. DNN applications such as healthcare~\cite{krumholz2016data, tan2020multimodal, suresh2020deep}, chatbot~\cite{openai2023gpt4}, and recommendation~\cite{naumov2019deep} jointly process heterogeneous data types (e.g., text, images, and tabular data). Specifically, the rise of generalist AI models, such as GPT-4o~\cite{gpt-4o}, Chameleon~\cite{team2024chameleon}, and Gato~\cite{reed2022generalist}, further underscores the need for efficient parallel handling of diverse data modalities. DNNs employed therein are designed to feature multiple branches, which are computationally independent and thus can be executed concurrently.
But existing DNN systems with \spp first linearize the computation graph of a DNN to construct the stages of a sequential pipeline and process these stages sequentially, falling short in harnessing the opportunity to blend such branch-level parallelism with pipeline parallelism.

Second, pipeline depth (i.e., number of sequential stages in \spp) is unduly increased by missing parallelism opportunities that arise from inherent DNN structures (e.g., parallel branches). Under an alternative arrangement in which some pipeline stages are parallelized by exploiting such structures, the number of sequential stages a micro-batch traverses in a forward (or backward) pass can be smaller. That is, the elongated pipeline formed by \spp unduly increases pipeline depth, which in turn increases the number of in-flight micro-batches to manage. This imposes a higher burden of managing memory, especially for early stages in the pipeline. Recall that the tight memory constraint in training large DNNs is a primary reason to apply pipeline parallelism. Thus, it is critical to curb the heightened memory requirement.

Third, today's devices employed for DNN training (e.g., GPUs) have high parallel-computing capabilities, requiring a large amount of training samples to be fetched to achieve peak performance. The increased memory consumption that results from applying \spp impedes doing so. As a consequence, devices perform computations at an operational intensity lower than their desired capacity, resulting in suboptimal training performance.

\begin{figure*}
    \centering
    \includegraphics[width=0.95\textwidth]{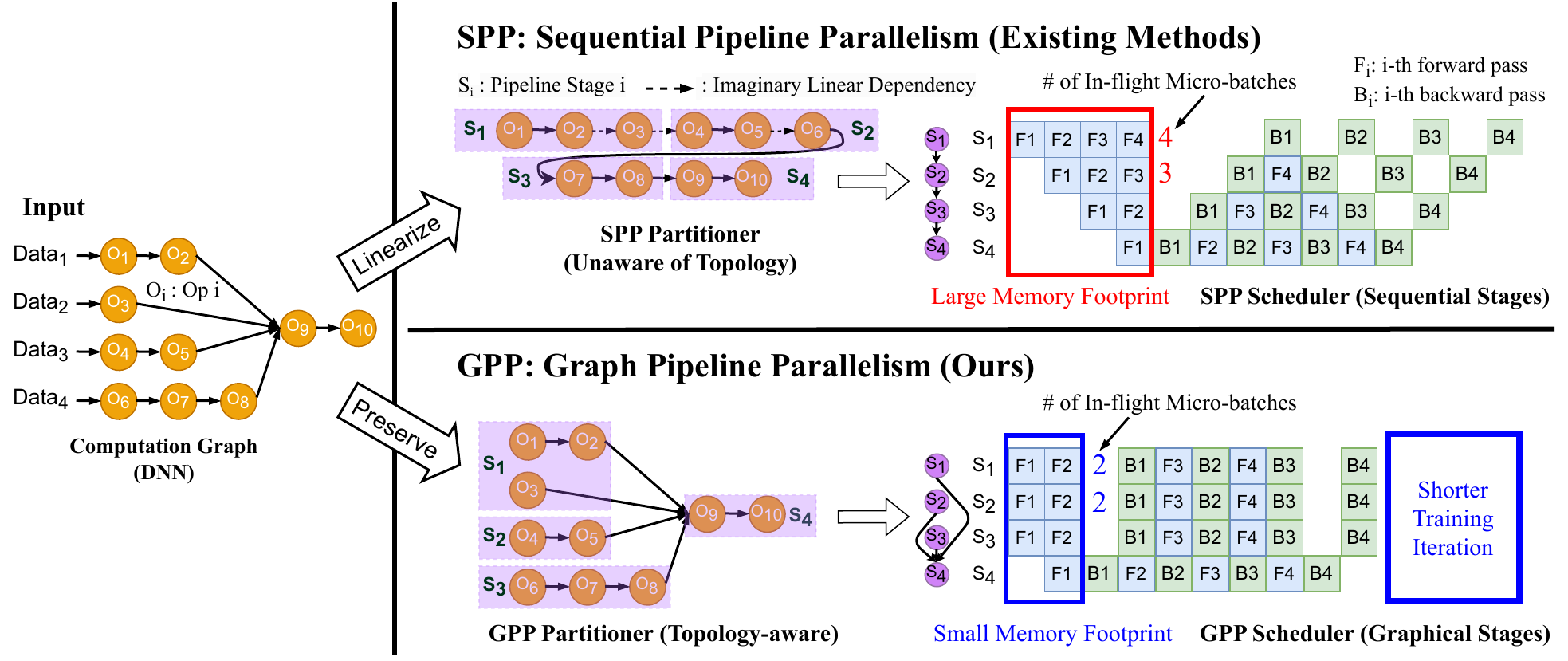}
    \caption{A high-level comparison between existing (\spp) and our (\gpp) approaches. \spp (top) produces sequential pipeline stages that miss the opportunity of parallelizing the branches in the DNN. In contrast, \gpp (bottom) generates graphical pipeline stages that enable {\em parallel execution} of the branches. This leads to lower training iteration time (i.e., higher training throughput) and smaller memory footprint in pipeline-parallel DNN training.}
    \label{fig:gpp_example}
\end{figure*}

\paragraph{Our approach.} To address the above challenges, we introduce {\em graph pipeline parallelism} (\gpp), that enables performant and scalable DNN training. \Cref{fig:gpp_example} highlights the key difference between \gpp and \spp. Instead of enforcing a strictly sequential execution order of pipeline stages, \gpp allows partitioning a DNN into stages whose dependencies are identified by a directed acyclic graph. \gpp includes \spp as a special case and can preserve the inherent topology of the DNN during stage partitioning.
As a result, \gpp enables concurrent execution of computationally-independent components, resulting in reduced memory requirement and improved GPU performance compared to \spp. 

\gpp involves a significantly larger and more complicated search space of parallelization strategies compared to the \spp strategies considered by existing DNN systems. Discovering \gpp strategies with superior performance over existing \spp baselines requires weighing subtle trade-offs between pipeline depth, memory consumption, and micro-batch schedule. To unleash the power of \gpp, we develop \ours, a system that automatically discovers efficient \gpp strategies to enable performant and scalable DNN training. \ours includes three key components. First, a {\em pipeline stage partitioner} automatically determines how to partition the operators of a DNN into a graph of stages, while balancing the computational load among these stages and minimizing inter-stage communication. Second, a {\em static micro-batch scheduler} schedules the forward and backward passes of different micro-batches within a mini-batch to minimize the peak GPU memory requirement of a \gpp strategy. The stage partitioner and micro-batch scheduler jointly partition a DNN into stages and determine the micro-batch schedules for each stage. Finally, a {\em distributed runtime} uses the discovered \gpp strategy to enable performant and scalable DNN training.

Through experiments on three multi-branch DNNs (e.g., Multi-Modal Transformer~\cite{vaswani2017attention, wang2023large, radford2021learning, openai2023gpt4, ramesh2021zero, jia2021scaling}, DLRM \cite{naumov2019deep}, and CANDLE-Uno \cite{2018candleuno}), we show that \ours can achieve up to 1.6$\times$ training throughput improvements over existing pipeline-parallel systems such as PipeDream~\cite{narayanan2019pipedream} and Piper~\cite{tarnawski2021piper}. 
\ours also reduces the search time
by 9-21$\times$ compared to PipeDream and Piper.

To summarize, we make the following contributions:
\begin{itemize}
    \item We introduce graph pipeline parallelism, a new parallelization scheme that promotes concurrent stage execution, reduces memory requirement, and improves GPU utilization compared to existing \spp schemes.
    \item We design algorithms to partition a DNN into a graph of stages and schedule micro-batches for these stages, which jointly discover efficient \gpp strategies.
    \item We develop \ours, a distributed runtime that enables fast and scalable DNN training with \gpp.
\end{itemize}

\makeatletter
\DeclareRobustCommand{\cev}[1]{%
  {\mathpalette\do@cev{#1}}%
}
\newcommand{\do@cev}[2]{%
  \vbox{\offinterlineskip
    \sbox\z@{$\m@th#1 x$}%
    \ialign{##\cr
      \hidewidth\reflectbox{$\m@th#1\vec{}\mkern4mu$}\hidewidth\cr
      \noalign{\kern-\ht\z@}
      $\m@th#1#2$\cr
    }%
  }%
}

\newcommand{\sche}[2]{{#1}_{#2}}
\newcommand{\gpforward}[3]{\vec{{#1}}_{#2}^{(#3)}}
\newcommand{\gpbackward}[3]{\cev{{#1}}_{#2}^{(#3)}}

\section{Graph Pipeline Parallelism}
\label{sec:gpp}

\Cref{fig:gpp_example} describe the key differences between sequential pipeline parallelism (\spp) employed by existing DNN systems~\cite{tarnawski2021piper, narayanan2021memory} and graph pipeline parallelism (\gpp). 
Given a DNN and a set of devices, \spp and \gpp produce strategies with different partitioning of stages and pipeline schedules.

\textbf{Concurrent execution of stages.} \spp \emph{linearizes} all operators
of a DNN while preserving data dependencies between these operators, and then partitions the linearized DNN into a sequence of pipeline stages. As a result, each stage has at most one predecessor and one successor. The execution order of these stages is thus {\em strictly sequential}.

In contrast, \gpp preserves the topology of a DNN when partitioning it into pipeline stages. To avoid circular dependencies between pipeline stages, the relationships between these stages form a {\em directed acyclic graph}. The execution order of the stages can be thus more general compared to \spp. This topology-aware partitioning and pipeline stage execution provides \gpp a clear advantage: (potentially) concurrent execution of stages that are computationally-independent.

For the \gpp strategy in \Cref{fig:gpp_example}, the three stages $S_1$, $S_2$, and $S_3$ are computationally-independent. Accordingly, the forward and backward passes of the three stages can be executed concurrently. However, in the \spp strategy, the two stages $S_2$ and $S_3$ are partitioned such that they have a sequential data dependency (due to the dependency between operator $o_6$ in $S_2$ and operator $o_7$ in $S_3$) since the \spp partitioner does not consider the topology of the DNN and fails to exploit it.
Moreover, while the two stages $S_1$ and $S_2$ in the \spp strategy should be computationally-independent according to the original DNN, the \spp scheduler executes the forward and backward passes of these two stages sequentially. This is because new data dependencies are imposed between them when linearizing the operators of a DNN to construct a sequential pipeline.

This distinction directly leads to a performance gap.
Specifically, both \spp and \gpp involve a {\em warm-up phase} during which micro-batches are injected into the pipeline until all stages can perform work concurrently.
However, as shown in \Cref{fig:gpp_example}, the warm-up phase of \gpp (i.e., 2) is shorter than that of \spp (i.e., 4). This performance improvement also applies to the {\em cool-down phase} during which in-flight micro-batches are resolved. As a result, \gpp achieves a shorter per-iteration training time (hence, a higher throughput) than \spp. The topology-aware stage partitioning and scheduling of \gpp address the first shortcoming of \spp (\S\ref{sec:intro}).

\textbf{Reduced memory requirement.} 
There is a close relationship between the memory requirement of a pipeline-parallel strategy and its pipeline {\em depth}, which is defined as the diameter of its stage graph. As mentioned earlier, the depth of the pipeline becomes excessively extended due to overlooked opportunities for parallelism within DNN architectures, such as parallel branches. This results in a higher memory footprint compared to that of pipeline stages that are parallelized by taking advantage of these structures.

In Figure~\ref{fig:gpp_example}, \gpp and \spp have a pipeline depth of 2 and 4, respectively. As a result, the first stage with the highest activation memory pressure needs to store the forward pass results for 2 micro-batches in \gpp and those for 4 micro-batches in \spp. All else being equal (i.e., an identical model partition by both), \gpp has a lower total memory footprint than \spp.
Note that memory saving is likely to grow as model size grows since a bigger model with deeper pipeline depth requires larger number of in-flight micro-batches (especially for early stages). The activation memory saving by \gpp addresses the second shortcoming of \spp in \S\ref{sec:intro}.

\textbf{Improved GPU utilization.} Devices employed in DNN training (e.g., GPUs) are designed to parallelize DNN computation of a micro-batch efficiently. Thus, larger micro-batches (i.e., more training samples within a micro-batch) can improve the operational intensity, thus GPU utilization of DNN operators. Note that larger micro-batches lead to reduced numbers of micro-batches, which in turn increases the warm-up and cool-down time of pipeline that \gpp can significantly reduce.
For simplicity of presentation, \Cref{fig:gpp_example} assumes that the same micro-batch size is used by \gpp and \spp. However, a lower device memory requirement of \gpp over \spp allows integrating more training samples in a micro-batch, which increases the operational intensity and overall GPU utilization, and therefore further reduces the per-iteration training time.
We evaluate this aspect in more detail in \S\ref{sec:evaluation}.

\section{Problem Formulation}
\label{sec:setup}

In this section, we formulate the problem of devising a \gpp strategy for distributed DNN training. Compared to existing works~\cite{tarnawski2021piper, narayanan2021memory, zheng2022alpa}, we further generalize the formulation to support graphical pipeline stages with fine-grained per-stage micro-batch size and schedules. As input, we are given (a) a computation graph $\m{G}_C = (\m{V}_C, \m{E}_C)$ that represents the neural architecture of a DNN model, (b) a mini-batch size $B$, and (c) a device topology graph $\m{D} = (\m{V}_D, \m{E}_D)$ where each node $v \in \m{V}_D$ represents a device with memory budget $M_v$ and each edge $e \in \m{E}_D$ represents a communication link with bandwidth $C_e$ between the two adjacent devices.

As output, we generate a {\em pipeline stage graph} $\m{G}_S = (\m{V}_S, \m{E}_S)$ that optimizes the performance metric of interest. In this work, we limit the scope to strategies that combine pipeline-parallel and data-parallel techniques, and aim to minimize the Time-Per-Sample (TPS) of the bottleneck pipeline stage since the pipeline throughput performance hinges upon the straggler stage.
The stage graph $\m{G}_S = (\m{V}_S, \m{E}_S)$ is a directed acyclic graph (DAG), where each node $S_i \in \m{V}_S$ specifies a pipeline stage and each directed edge $(S_i, S_j) \in \m{E}_S$ indicates that stage $S_i$ must precede $S_j$ for forward passes and that $S_j$ must precede $S_i$ for backward passes.


The goal is to solve the min-max optimization problem:
\begin{align}
\min \quad & \max_{S_i \in \m{V}_S} \mathsf{TPS}(S_i ; \m{G}_C, B, \m{D}) \label{opt:objective} \\
\textrm{s.t.} \quad & \max_{S_i \in \m{V}_S} \mathsf{DeviceMemoryUsage}(S_i ; \m{G}_C, B, \m{D}) \leq M. \label{opt:constraint}
\end{align}

Formally, \gpp devises a strategy $\m{G}_S$ as follows. We define $S_i \in \m{V}_S$ in further detail as a four-element tuple: $S_i = \langle \m{G}_i, b_i, \m{D}_i, \Pi_i \rangle$:
\begin{enumerate}
    \item $\m{G}_i$ represents a subgraph of $\m{G}_C$,
    \item $b_i$ is the micro-batch size of $S_i$ (i.e., there are $B/b_i$ micro-batches for each mini-batch),
    \item $\m{D}_i$ is a set of devices allocated to process the forward and backward passes of $S_i$ (we apply data parallelism within $S_i$ if $|\m{D}_i| > 1$), and
    \item $\Pi_i$ is a micro-batch schedule that specifies the order in which the $B/b_i$ forward and $B/b_i$ backward passes are processed. We use $\mathsf{fw}^i_j$ (or $\mathsf{bw}^i_j$) to denote the forward (or backward) pass of the $j$-th micro-batch for $S_i$.
\end{enumerate}

$\m{G}_S$ is a valid \gpp strategy if and only if the memory constraint (Equation~\ref{opt:constraint}) and all following conditions are met:
\begin{itemize}
    \item[\it C1.] $\m{G}_i$ is a convex subgraph of $\m{G}_C$
    , and $\m{G}_1, \dots ,\m{G}_{|\m{V}_s|}$ form a partition of $\m{G}_C$.
    \item [\it C2.] If there exists $(u, v) \in \m{E}_C$ such that $u \in \m{G}_i$ and $v \in \m{G}_j$, then $(S_i, S_j) \in \m{E}_S$.
    \item[\it C3.] $\m{D}_1,\dots,\m{D}_{|\m{V}_s|}$ form a partition of $\m{D}$.
    \item[\it C4.] For each micro-batch schedule $\Pi_i$, $\mathsf{fw}_{k}^i$ precedes $\mathsf{fw}_{k+1}^i$, $\mathsf{bw}_{k}^i$ precedes $\mathsf{bw}_{k+1}^i$, and $\mathsf{fw}_k^i$ precedes $\mathsf{bw}_k^i$.
\end{itemize}


In words, \textit{C1} mandates that all operators be covered by stages that do not overlap with each other, and \textit{C2} mandates that a strict sequential execution order between two stages be established if according to the computation graph there exists a data dependency between two operators each in either of the stages. \textit{C3} ensures that at least one device is allocated to every stage. \textit{C4} dictates the orderings of forward and backward passes.

\section{System Overview}
\label{sec:overview}

\begin{figure}[t!]
    \centering
    \includegraphics[width=\columnwidth]{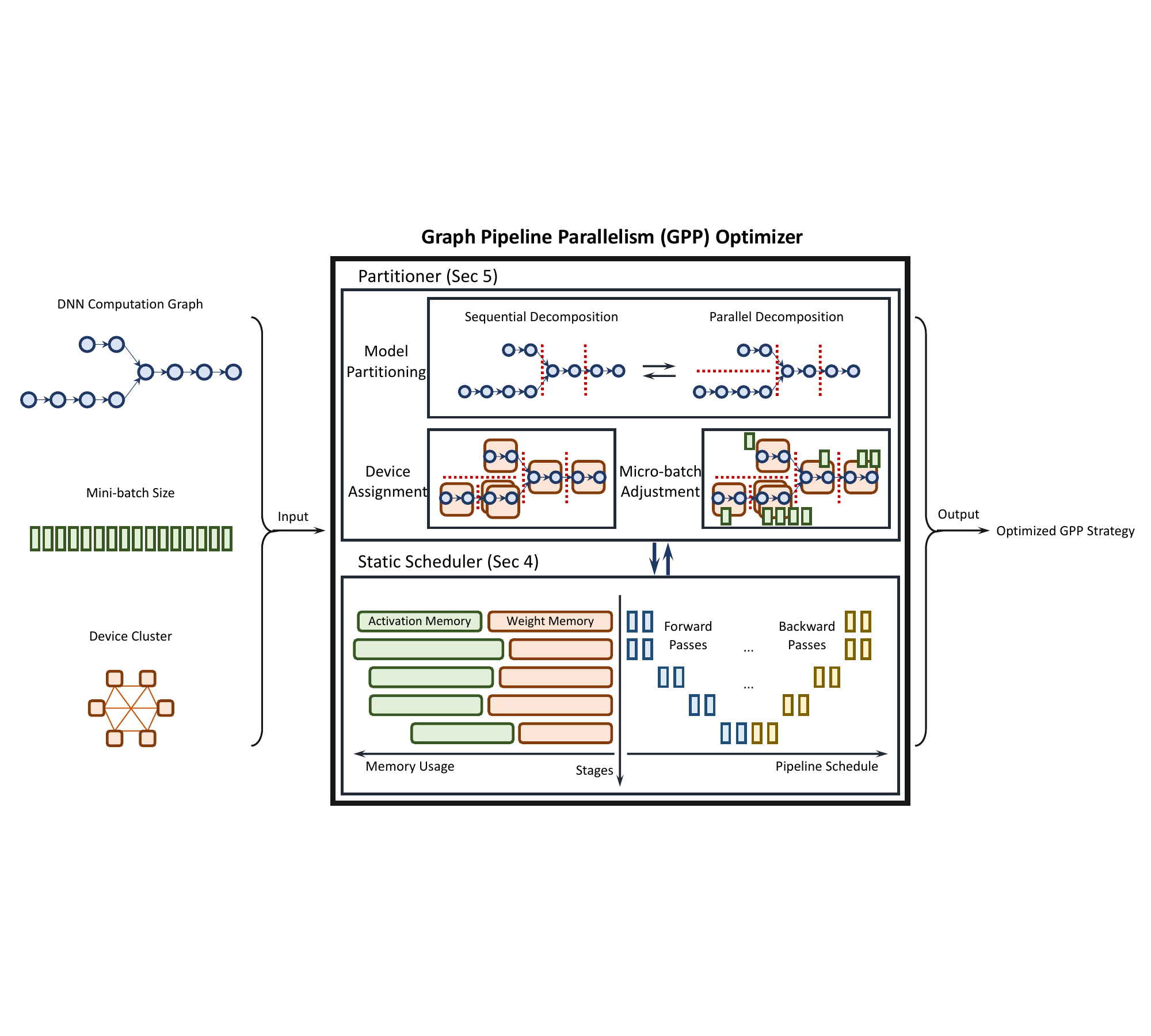}
    \caption{Overview of \ours. It consists of a pipeline stage partitioner and a micro-batch scheduler. Given a DNN computation graph, mini-batch size, and device configuration, they interact with each other to produce an optimized \gpp training strategy as output. The output can be launched on the distributed runtime framework we also develop to execute it and evaluate its real-world performance.}
    \label{fig:system-overview}
\end{figure}

\Cref{fig:system-overview} illustrates an overview of \ours, a system that accelerates distributed DNN training at scale using \gpp. Taking as input (a) the computation graph of a DNN, (b) mini-batch size, and (c) the topology of assigned GPUs, \ours produces an optimized \gpp strategy for parallel DNN training.
\ours includes three key components: a pipeline stage partitioner, a static micro-batch scheduler, and a distributed runtime. The first two components jointly discover a high-performance \gpp strategy for a given DNN model, mini-batch size, and assigned devices, which will be executed by the distributed runtime.

\textbf{Pipeline stage partitioner.} 
The partitioner performs three tasks. First, it partitions a DNN, aimed at achieving an effective distribution of workloads across stages. It examines the amount of computation and communication needs associated with the operators in each stage. Importantly, it \emph{leverages the inherent topology} of the DNN at hand in order to exploit \emph{concurrent} execution opportunities. To this end, it performs a sequence of series-parallel decompositions of the given DNN. Second, it adjusts the micro-batch size for each stage. This fine-grained adjustment aims to exploit heterogeneous compute efficiencies of different types of operators. Finally, it determines how many devices to assign to each stage to achieve an effective allocation of resources. Note that all three functions are jointly performed, as no one function is independent of the others. We provide further details in \S\ref{sec:partitioner}.

\textbf{Static micro-batch scheduler.} The scheduler performs two tasks. First, it optimizes micro-batch schedules for forward and backward passes while ensuring the integrity of distributed DNN training. This involves examining both intra- and inter-stage data dependencies between the passes (see {\it C4} in \S\ref{sec:setup}). Next, it checks if the memory usage that results from the schedule is within the given device memory constraint (see Equation~\ref{opt:constraint}). Memory usage is closely related to the numbers of in-flight micro-batches of a stage, which can be computed based on the schedule of the forward and backward passes of the stage. \S\ref{sec:scheduler} provides further details.


\textbf{Distributed runtime framework.} We develop a distributed DNN runtime system that executes \gpp training strategies generated by the optimizer of \ours. Using the distributed runtime as the testbed, we compare the performance of the generated \gpp strategies against existing \spp strategies for various DNNs. We provide details in \S\ref{sec:evaluation}.





\section{Pipeline Stage Partitioner}
\label{sec:partitioner}


\begin{algorithm} [th!]
	\small
	\caption{Pipeline stage partitioner.} \label{alg:faster-partitioner-sp-dp}
	\textbf{Input}: Computation graph $\m{G}_C$, number of devices $|\m{V}_D|$\\ 
	\textbf{Output}: Optimized stage graph $\m{G}_S$\\
	\begin{algorithmic}[1]
        \State // $\mathsf{MAXTPS}$: safe upper-bound for TPS of bottleneck stage.
        \State $t_{l} = 0$, $t_{r} = \mathsf{MAXTPS}$, $\m{G}_{S} = \varnothing$
        \label{alg:binary_start}
        \While{$t_r - t_l > \epsilon$}
            \State $t_m = (t_l + t_r)/2$
            \State $\m{G}_{S}^{best} =$ \Call{SearchStageGraph}{$\m{G}_C, |\m{V}_D|, t_m, B$}
            \If{$\m{G}_{S}^{best} == \varnothing$}
                \State $t_l = t_m$
            \Else
                \State $t_r = t_m$
                \State $\m{G}_{S} = \m{G}_{S}^{best}$
            \EndIf
        \EndWhile
        
        \State \textbf{return} $\m{G}_{S}$ \label{alg:binary_end}
        \State

        \Function{SearchStageGraph}{$\m{G}_C, |\m{V}_D|, t_m, B$} \label{alg:dp_val_start}
            \State // $C$ is a set of candidate schedule configurations ($c$)
    		\For {$c \in C$}
        	    \State // $c_0$: dummy schedule configuration
                    \State $\m{G}_{S}^{new}$ = \Call{DP}{$\m{G}_C, c_0, c, |\m{V}_D|, t_m$}
                    \State // \Call{PickBetter}{$\cdot$} picks one with less memory
                    \State $\m{G}_{S}^{best} = $ \Call{PickBetter}{$\m{G}_{S}^{best}, \m{G}_{S}^{new}$}
            \EndFor
            \State \textbf{return} $\m{G}_{S}^{best}$ \label{alg:dp_val_end}
        \EndFunction
        
        \State
        \State // Dynamic Programming (DP) Partitioner
	    \Function{DP}{$\m{G}, c_f, c_b, d, t_{max}$} \label{alg:dp_start}
                \If {this DP state has been visited}
                    \State \Return corresponding $\m{G}_{S}^{best}$ to this DP state
                \EndIf
                \State // Consider a given DP state as a SINGLE stage
                \State $\m{G}_{S}^{best} = \varnothing$
	        \If{\Call{EstimateTPS}{$\m{G}, c_f, c_b, d$}$\le t_{max}$} \label{dp-single}
                    \State // Optimize schedule via Algorithm~\ref{alg:microbatch-scheduling}
                    \State $\Pi_{opt}$ = \Call{ScheduleStage}{$\m{G}, c_f, c_b, d$}
                    \State $\m{G}_{S}^{best} = $ \Call{StageGraph}{$\m{G}, \Pi_{opt}, d$}
    		\EndIf
                \State // Decompose a given DP state into two stages
	        \If {$\m{G}$ can be decomposed in series} \label{dp-serial}
    			\For {$(\m{G}_1, \m{G}_2) \in$ \Call{SeriesDecompose}{$\m{G}$}}
    			    \For {$d_2 \gets 1$ to $d-1$}
                            \State $d_1 = d - d_2$
                            \For {$c_m \in C$}
                    	    \State $\m{G}_{S_2}^{new} = $ \Call{DP}{$\m{G}_2, c_m, c_b, d_2, t_{max}$}
                                \State Update $i_m$ based on $\m{G}_{S_2}^{new}$
                        	\State $\m{G}_{S_1}^{new} =$\Call{DP}{$\m{G}_1, c_f, c_m, d_1, t_{max}$}
                            \EndFor
                        \EndFor
                    \EndFor
			\ElsIf {$\m{G}$ can be decomposed in parallel} \label{dp-parallel}
	               \For {$(\m{G}_1, \m{G}_2) \in$ \Call{ParallelDecompose}{$\m{G}$}}
    			    \For {$d_1 \gets 1$ to $d-1$}
                        \State $d_2 = d - d_1$
                	    \State $\m{G}_{S_1}^{new} = $ \Call{DP}{$\m{G}_1, c_f, c_b, d_1, t_{max}$}
                	    \State $\m{G}_{S_2}^{new}  = $ \Call{DP}{$\m{G}_2, c_f, c_b, d_2, t_{max}$}
                        \EndFor
                    \EndFor
            \EndIf
            \State $\m{G}_{S}^{best} = $ \Call{PickBetter}{$\m{G}_{S}^{best}, \m{G}_{S_1}^{new} \cup \m{G}_{S_2}^{new}$}
	    \State \Return $\m{G}_{S}^{best}$ \label{alg:dp_end}
	    \EndFunction

	\end{algorithmic}
\end{algorithm}

The pipeline stage partitioner of \ours aims to minimize Time-Per-Sample (TPS) of the bottleneck pipeline stage as described in \S\ref{sec:setup}. It takes as input a DNN computation graph $\m{G}_C$, a mini-batch size $B$, and a device topology graph $\m{G}_D$, and generates an optimized stage graph $\m{G}_S$ by searching over different model partitions, device assignments, and micro-batch sizes simultaneously. A key challenge we must address is the large and complex search space of potential \gpp strategies. To reduce the complexity of the search task, we employ a binary search method combined with series-parallel decomposition and dynamic programming. We next describes these three components.

\textbf{Binary search.} Given the large search space of potential solutions, \ours does not attempt to directly find an optimal solution. Instead, \ours employs binary search to iteratively narrow down the target performance range and examines whether there exist valid solutions within the range. By  iteratively reducing the range, \ours discovers solutions arbitrarily close to an optimal one, and thus there is little difference in performance for practical purposes. Lines \ref{alg:binary_start}--\ref{alg:binary_end} of \Cref{alg:faster-partitioner-sp-dp} shows \ours's binary search process.

\textbf{Series-parallel decomposition.} Since most DNNs structurally reflect series-parallel graphs~\cite{series-parallel, unger2022unity}, \ours applies series-parallel decomposition to an input graph $\m{G}_C$ in order to decompose it into smaller, manageable subgraphs, and perform model partitioning, device allocation, and task scheduling for each subgraph. In the unusual cases where a DNN does not possess such a structural property, \ours bypasses this issue by converting the DNN to an arithmetically identical one whose structure is a series-parallel graph.




\textbf{Dynamic programming (DP).} \ours adopts a dynamic programming algorithm where the value of each DP state indicates the existence of a strategy achieving a throughput within a target range (Lines \ref{alg:dp_val_start}--\ref{alg:dp_val_end} of \Cref{alg:faster-partitioner-sp-dp}).
At each DP level, \ours applies series-parallel decompositions to split an input graph (say $\m{G}$) into two new subgraphs (say $\m{G}_1, \m{G}_2$), each of which serves as the input computation graph of a new DP subproblem at one DP level below.
\ours recursively solves the DP subproblems to construct a solution of the original problem where the input computation graph is $\m{G}_C$ (Lines \ref{alg:dp_start}--\ref{alg:dp_end} of \Cref{alg:faster-partitioner-sp-dp}).


\textbf{DP subproblem.} We ensure that each DP subproblem maintains a certain structure (i.e., having a unique pair of source and sink nodes and a subgraph $\m{G}$ comprised of them).
The input to a DP subproblem includes a computation graph $\m{G} \subseteq \m{G}_C$, the number of devices $d$, and some schedule-related information for its predecessor and successor stages, which we furnish by enumeration if not available.

The solution of a DP subproblem involves devising a training strategy such that (1) the number of in-flight micro-batches for the source stage (i.e., the pipeline stage that includes the source node) is minimized; and (2) the Time-Per-Sample (TPSes) for all stages do not exceed the target TPS range. These results are returned back to the parent DP subproblem at one DP level above where the results are gathered for the parent DP subproblem to produce its own.

We consider three cases in a DP subproblem:
\begin{itemize}
    \item \textit{Base case:} We consider the entire subgraph $\m{G}$ as a single stage and apply data parallelism with a data-parallel degree of $d$ (Line~\ref{dp-single} in Algorithm~\ref{alg:faster-partitioner-sp-dp}). We estimate TPS by profiling the execution time of each operator while extrapolating communication latency by affine functions. We check if the target TPS range is achievable with the memory constraint, and compute the number of in-flight micro-batches according to Algorithm~\ref{alg:microbatch-scheduling} (see \S\ref{sec:scheduler}). 
    \item \textit{Series decomposition:} We perform a series decomposition to create two subgraphs $\m{G}_1$ and $\m{G}_2$, where the sink node of $\m{G}_1$ coincides with the source node of $\m{G}_2$ (Line~\ref{dp-serial} in Algorithm~\ref{alg:faster-partitioner-sp-dp}).
    We first solve the subproblem associated with $\m{G}_2$. To do so, we enumerate all feasible schedules for the source node of $\m{G}_2$. We then solve the subproblem associated with $\m{G}_1$. 
    \item \textit{Parallel decomposition:} We perform a parallel decomposition to create $\m{G}_1$ and $\m{G}_2$, where $\m{G}_1$ and $\m{G}_2$ share the same source and sink nodes (Line~\ref{dp-parallel} in Algorithm~\ref{alg:faster-partitioner-sp-dp}). As there is no data dependency between these subgraphs, the pipelines can be executed in parallel. The subproblems associated with $\m{G}_1$ and $\m{G}_2$ may produce different optimal numbers of in-flight micro-batches for the shared source node. To ensure continuous pipelining, we take the larger number of in-flight micro-batches as the solution.
\end{itemize}

\begin{figure}[h!]
    \centering
    \includegraphics[width=\columnwidth]{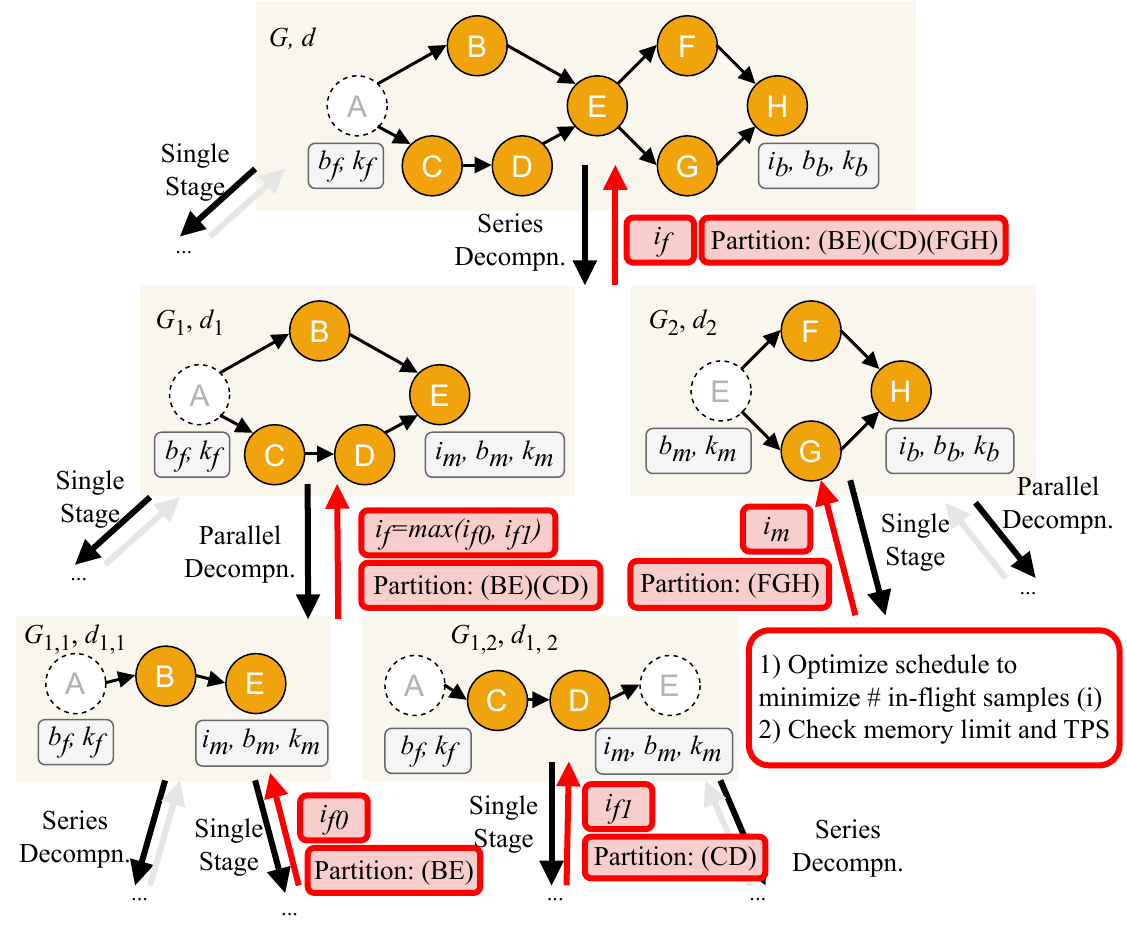}
    \caption{Pipeline stage partitioner performing series-parallel decompositions. Black arrows indicate subproblem formulations. Red arrows indicate solutions of subproblems.}
    \label{fig:dp}
\end{figure}

\textbf{Overall process.} Figure~\ref{fig:dp} visualizes the overall process. At the top, a DP subproblem is provided with its initial conditions: computation graph $\m{G}$, the number of available devices $d$, and the target TPS range $[0, t_{max}]$.
Suppose the number of in-flight micro-batches for the sink node is $i_b$, the micro-batch sizes for the source and sink nodes are $b_f$ and $b_b$, the stage containing the source node (i.e., source stage) uses the $k_f$F$k_f$B schedule, and the stage containing the sink node (i.e., sink stage) uses the $k_b$F$k_b$B schedule (we introduces \ours's micro-batch schedules in \S\ref{sec:scheduler}). These supposed conditions comprise a schedule configuration denoted by $c := (i, b, k)$\footnote{Find the definition of $i, b, k$ in \Cref{sec:nfnbrule}} in Algorithm~\ref{alg:microbatch-scheduling}. They are either available as the results of some other DP subproblems solved previously, or furnished by enumeration.
The solution of this DP subproblem computes the smallest possible number of in-flight micro-batches for the source stage (i.e., $i_f$ in Figure~\ref{fig:dp}) that meets the target TPS range $[0, t_{max}]$.

\textbf{Time complexity.} We analyze the time complexity of the stage partitioner to gauge the impacts of design parameters. Let $N$ be the number of series-parallel subgraphs of $\mathcal{G}_C$, $\mathcal{B}$ be the set of possible micro-batch sizes, $\mathcal{D}$ be the set of possible data-parallel degrees. The maximal element of $\mathcal{B}$ is upper-bounded by $B$. We consider powers of 2 for micro-batch sizes (i.e., $|\mathcal{B}| < \log_2 B$). Likewise, the maximal element of $\mathcal{D}$ is upper-bounded by $|\mathcal{V}_D|$ and $|\mathcal{D}| < \log_2 |\mathcal{V}_D|$ holds.

The number of candidates for $\m{G}$ is $O(N)$, that for $c_f = (b_f, k_f)$ is $O(|\mathcal{B}|^2)$, that for $c_b = (i_b, b_b, k_b)$ is $O(B|\mathcal{B}|^2)$, and that for $d$ is $O(|\mathcal{D}|)$ in each DP subproblem. To compute a DP value, it takes $O(|\mathcal{D}||\mathcal{B}|^2)$ time for series decompositions and $O(|\mathcal{D}|)$ time for parallel decompositions. Therefore, the time complexity for a single DP run is $O(NB|\mathcal{B}|^6|\mathcal{D}|^2)$ and the overall time complexity is $O((\log \mathsf{MAXTPS}) NB|\mathcal{B}|^6|\mathcal{D}|^2) = O((\log \mathsf{MAXTPS}) NB(\log_2 B)^6(\log_2 |\mathcal{V}_D|)^2)$.



\begin{figure*}
    \centering
    \includegraphics[width=0.95\textwidth]{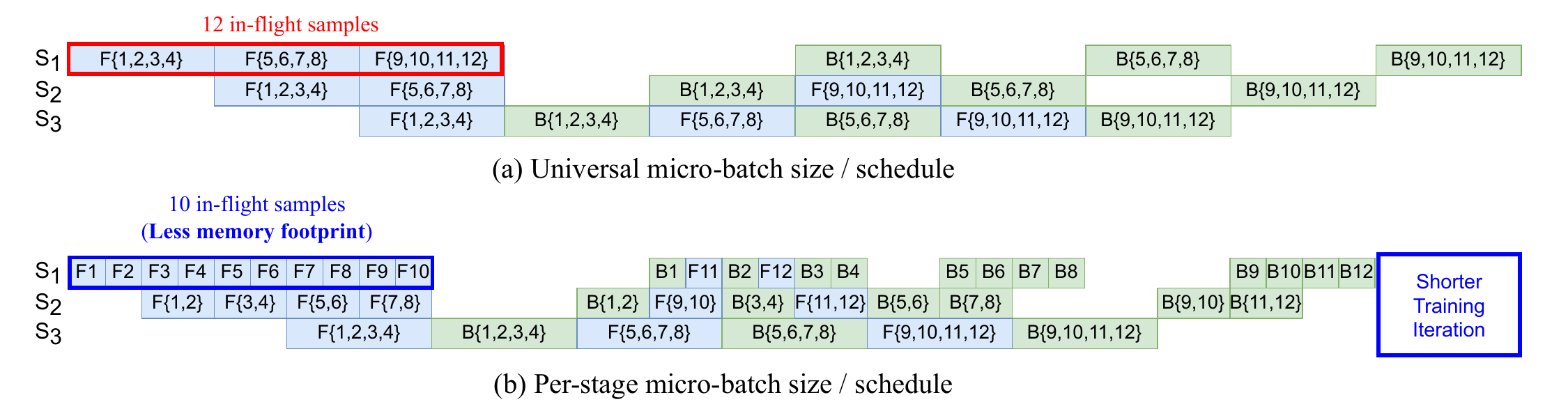}
    \caption{A comparison between universal and per-stage micro-batch size / schedule. F$\{i,j\}$, B$\{i,j\}$ indicate forward and backward passes for a micro-batch including samples $i$ and $j$. It showcases how per-stage micro-batch size and scheduling can save  memory footprint and training iteration time.}
    \label{fig:1f1b_nfnb}
\end{figure*}

\section{Static Micro-Batch Scheduler}
\label{sec:scheduler}

The static micro-batch scheduler of \ours optimizes micro-batch size and schedules to minimize training time and memory footprint. Specifically, we design our scheduler to address the unique challenges presented by graph-like data dependencies in \gpp pipeline stages. These dependencies make scheduling non-trivial unlike \spp case. For example, as shown in \Cref{fig:pp}, it is straightforward in \spp that we need to schedule $n + 1- i$ ($n$: the number of sequential stages) forward tasks at stage $i$ until we schedule a first backward task, assuming 1F1B schedule (i.e., forward pass for 1 micro-batch followed by backward pass for 1 micro-batch)\footnote{A schedule $\Pi_i$ is said to be $k$F$k$B when there exist $\ell$ and $k$ such that $\Pi_i$ starts with $\ell$ forward passes (for warm-up), alternates between $k$ backward and $k$ forward passes, and ends with $\ell$ backward passes (for cool-down).}. However, with \gpp, this simple equation does not hold anymore since there could be multiple stages following a single stage. Therefore, we need more generalized method to optimize schedules while meeting the graph-like data dependency between all forward and backward tasks.

We further generalize our scheduler so that it can support different micro-batch sizes and schedules over pipeline stages. This can be effective when running heterogeneous models (e.g., multi-modal models with different ideal micro-batch size over pipeline stages across different modalities).

\Cref{fig:1f1b_nfnb} illustrates how we can reduce training iteration time and memory footprint with per-stage micro-batch size and scheduling. Here, each pipeline stage has different smallest micro-batch size (i.e., 1, 2, 4 for $S_1, S_2, S_3$) achieving maximum compute efficiency. Using a fixed micro-batch size of 4 across all stages maximizes compute efficiency. But, the drawback is long GPU idling during warm-up and cool-down, and large memory footprint, i.e., 12 in-flight micro-batches for stage 1. However, by tailoring the micro-batch size and scheduling to each stage, we reduce this to 10 in-flight micro-batches for stage 1 while maintaining maximum compute efficiency. This also shortens the training iteration because stages can be scheduled earlier with smaller micro-batch sizes. This benefit usually grows with more pipeline stages.

\begin{algorithm}[th!]
	\small
	\caption{Static micro-batch scheduler.} \label{alg:microbatch-scheduling}
	\textbf{Input}: Model partition $\m{G}$, initial current and next stage schedule configurations $c_f, c_b$, number of devices $d$ \\
	\textbf{Output}: Optimized schedule $\Pi_{opt}$\\
	\begin{algorithmic}[1]
            \Function{ScheduleStage}{$\m{G}, c_f, c_b, d$}
                \State // Optimize schedule by minimizing number of 
                \State // in-flight micro-batches 
                \State // while respecting data dependencies
                \State $i_f = $ \Call{ComputeInFlight}{$k_f, b_f, k_b, b_b, i_b$}
                \State $c_{opt} = (i_f, k_f, b_f)$
                \If {$c_{opt}$ violates device memory constraint}
                    \State $c_{opt} = \varnothing$ // Invalidate schedule $c_{opt}$
                \EndIf
                \State $\Pi_{opt} \gets$ \Call{ScheduleTask}{$c_{opt}$}
                \State \Return $\Pi_{opt}$ 
            \EndFunction
	\end{algorithmic}
\end{algorithm}

To support 1) graph-like stage dependency and 2) per-stage micro-batch size and schedule, this is how we design our scheduler (\Cref{alg:microbatch-scheduling}). It takes as input (1) a configuration of model partition $\m{G}$, (2) current and next stage schedule configurations $c_f, c_b$,  and (3) the number of devices $d$ from the pipeline stage partitioner, and produces an optimized micro-batch schedule $\Pi_{opt}$ for a given stage configuration. As in Figure~\ref{fig:system-overview}, the input is fed by the stage partitioner, and the output is returned back to the stage partitioner to form a stage graph with an optimized micro-batch schedule.



\ours's pipeline stage partitioner (\Cref{alg:faster-partitioner-sp-dp}) first calls \Cref{alg:microbatch-scheduling} to discover an optimized micro-batch schedule for the last stage. It then traces back all directed edges $(S_i, S_j) \in \m{E}_S$ of the stage graph $\m{G}_S$ in the reverse direction and determines a schedule for each stage $S_i$ until a schedule for the first stage is determined. The reason for backward traversal is that computing the activation memory usage, and thus the total usage, for a stage $S_i$ requires complete schedule information of its subsequent stages $S_j$.

$\textproc{ComputeInFlight}(\cdot)$ and $\textproc{ScheduleTask}(\cdot)$ are two key functions in \Cref{alg:microbatch-scheduling}. First, $\textproc{ComputeInFlight}(\cdot)$ is a key subroutine to optimize schedule by effectively minimizing the number of in-flight micro-batches for a given stage without increasing per-iteration training time. To take graphical stage dependency (i.e., multiple stages following one) into account, it factors in all following stages to decide the minimum number of in-flight micro-batches. It also accounts for scheduling constraints posed by micro-batch size gap between subsequent stages. For instance, in \Cref{fig:1f1b_nfnb}, $S_2$ needs two micro-batches to be processed from $S_1$ to process a single micro-batch.  \Cref{sec:nfnbrule} explains the detail of $\textproc{ComputeInFlight}(\cdot)$ computes the minimal number of in-flight micro-batches.

\begin{figure*}[th!]
    \centering
    \subfigure[Multi-Modal Transformer (MMT)]{\includegraphics[width=0.32\textwidth]{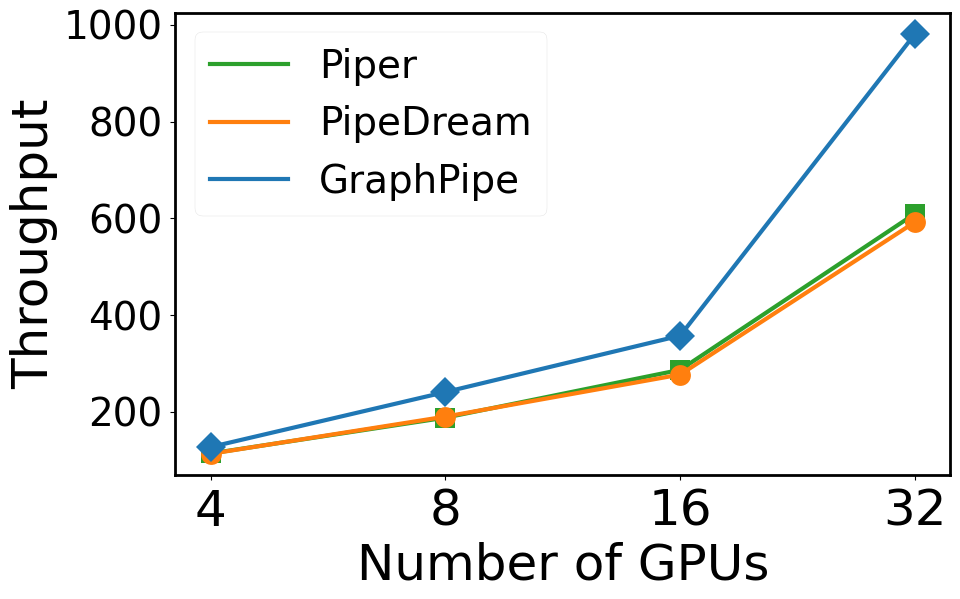}\label{e2e-perf-MMT}}
    \subfigure[DLRM]{\includegraphics[width=0.32\textwidth]{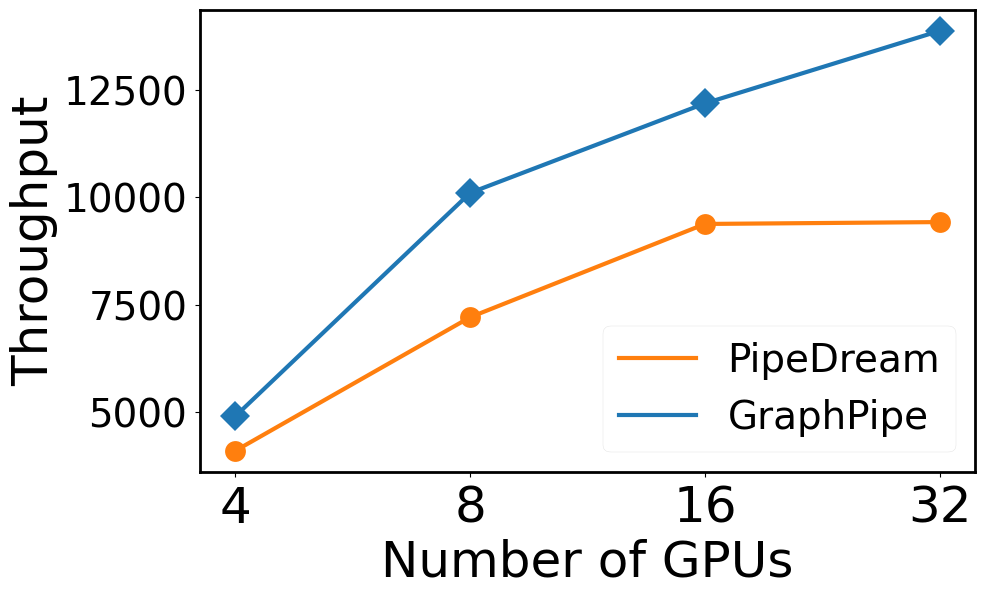}\label{e2e-perf-DLRM}}
    \subfigure[CANDLE-Uno]{\includegraphics[width=0.32\textwidth]{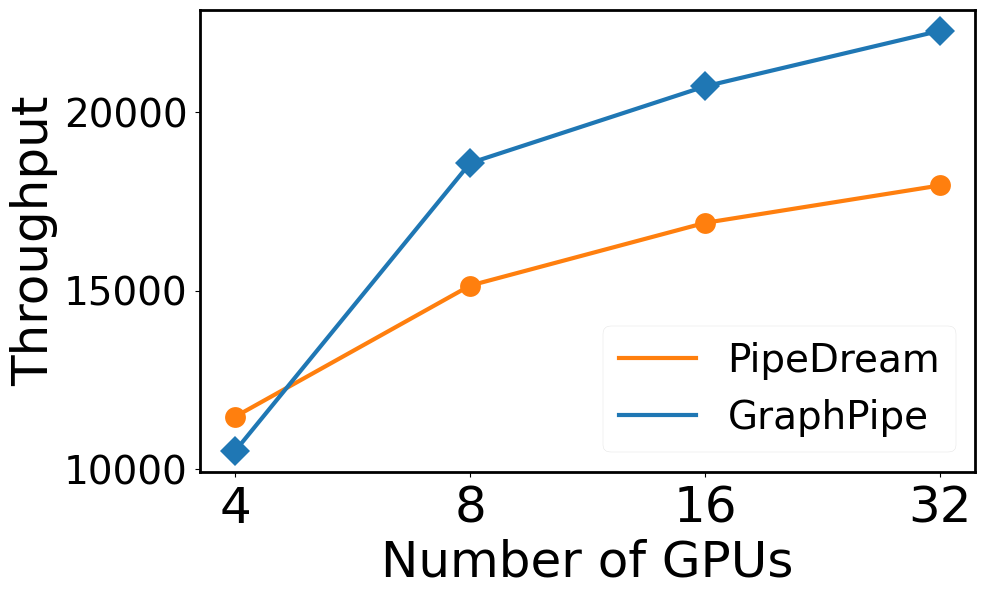}\label{e2e-perf-CANDLE-Uno}}
    \caption{End-to-end performance evaluation. \ours outperforms both \pd~\cite{narayanan2021memory} and Piper~\cite{tarnawski2021piper} in three different models: Multi-modal Transformer-based model \cite{radford2021learning}, DLRM \cite{naumov2019deep}, and CANDLE-Uno \cite{2018candleuno} at all but one GPU count configurations tested. Missing data points indicate that no training strategy can be found within reasonable timeframes.}
    \label{fig:e2e-plot}
\end{figure*}

Second, $\textproc{ScheduleTask}(\cdot)$ produces an optimized schedule of forward and backward passes with optimized schedule configuration ($c_{opt}$). It adopts greedy scheduling that schedules backward pass as early as possible.
It reduces both memory consumption and training iteration time since it quickly resolves the corresponding in-flight forward pass.

\ours uses the following default schedule configurations: 1) synchronous 1F1B schedule~\cite{narayanan2021memory} adjusted to support graph-like dependencies and 2) the same micro-batch size across stages. The synchronous 1F1B avoids gradient staleness with the same pipeline latency and lower activation memory footprint in comparison to alternatives (e.g., GPipe~\cite{huang2019gpipe}). Furthermore, except for some corner cases, we observe that performance improvements from per-stage micro-batch sizes and $k$F$k$B schedules are incremental to justify the increased search times for models and device clusters we explored. Still, with \ours, users can choose to search over per-stage micro-batch sizes and $k$F$k$B schedules for more heterogeneous models and larger device clusters.

\section{Evaluation}
\label{sec:evaluation}

We develop \ours on top of FlexFlow~\cite{jia2019beyond}, a distributed multi-GPU runtime for DNN training. We adjusted FlexFlow’s runtime for parallel execution of graphical pipeline stages while introducing our own partitioner in \S\ref{sec:partitioner} and scheduler in \S\ref{sec:scheduler} to FlexFlow. 
We evaluate \ours on the Summit supercomputer~\cite{summit}. For each compute node of Summit, we use 2 IBM POWER9 CPUs and 4 NVIDIA V100 GPUs with 512GB of main memory. GPUs within a node are interconnected via NVLink while nodes are connected via Mellanox EDR 100Gb InfiniBand. We use the default schedule configurations of \ours mentioned in \Cref{sec:scheduler}.
Note that we omit error bars for our plots, as we observe marginal standard deviations (less than 3\%) for all results.

\textbf{DNNs.} We explore three multi-branch DNNs: Multi-Modal Transformer-based model (MMT)~\cite{vaswani2017attention, radford2021learning}, DLRM~\cite{naumov2019deep}, and CANDLE-Uno~\cite{2018candleuno}. Multi-Modal Transformer (MMT) is a backbone of most state-of-the-art multi-modal models~\cite{wang2023large, radford2021learning, openai2023gpt4, ramesh2021zero, jia2021scaling}. DLRM is a popular deep learning recommendation model for personalization and ads recommendation. CANDLE-Uno is a specialized model in the medical domain (i.e., precision medicine). We describe the detailed model configurations in \Cref{sec:model_config}. 
Despite different applications, all these models feature parallel branches, each processing a different type of data.

\subsection{End-to-End Evaluation}

\begin{table*}[th!]
    \centering
    \begin{tabular}{c|ccc|ccc|ccc} \toprule
        \multirow{2}{*}{\# GPUs} & \multicolumn{3}{c|}{MMT} & \multicolumn{3}{c|}{DLRM} &  \multicolumn{3}{c}{CANDLE-Uno} \\  \cmidrule{2-10}
            & Piper & \pd & Ours & Piper & \pd & Ours & Piper & \pd & Ours  \\ \midrule
         4  & 52.9 (440.5$\times$) & 2.57 (21.4$\times$) & 0.12 & \xmark & 6.39 (19.3$\times$) & 0.33 & \xmark & 3.84 (20.2$\times$) & 0.19 \\
         8  & 126 (165.7$\times$) & 11.9 (15.6$\times$) & 0.76 & \xmark & 31.3 (11.4$\times$) & 2.73 & \xmark & 17.0 (11.8$\times$) & 1.43 \\
         16  & 304 (101.3$\times$) & 44.3 (14.7$\times$) & 3.00 & \xmark & 131 (9.9$\times$) & 13.28 & \xmark & 66.10 (10.7$\times$) & 6.14 \\
         32  & 745 (73.7$\times$) & 151 (15.0$\times$) & 10.11 & \xmark & 505 (9.2$\times$) & 54.6 & \xmark & 234 (10.4$\times$) & 22.37 \\ \bottomrule
    \end{tabular}
    \caption{Solution search times (in seconds) for Piper, \pd, and Ours (\ours) on the Apple M1 Max; \xmark \ indicates search cannot be completed. Numbers in parentheses indicate the search time ratio of the algorithm to that of \ours.}
    \label{tab:search-time}
\end{table*}

We compare the training throughput of \ours with existing pipeline-parallel systems such as \pd~\cite{narayanan2021memory} and Piper~\cite{tarnawski2021piper}. We choose these two baselines since their combined search space encompasses all possible model partitions covered by other \spp approaches~\cite{fan2021dapple, zheng2022alpa, narayanan2019pipedream}. To be specific, PipeDream (with the operator granularity) basically covers the pipeline partitioning and scheduling strategies of all baseline \spp approaches~\cite{fan2021dapple, zheng2022alpa, narayanan2019pipedream} but Piper. They all (1) linearize DNNs by transforming their computation graphs into sequences of operators, (2) exhaust pipeline stage partition choices, and (3) employ 1F1B scheduling.

\Cref{fig:e2e-plot} showcase the results. We measure the training throughput (i.e., number of samples processed per second) as we increase the number of GPUs and mini-batch sizes. Note that Piper cannot generate training strategies for DLRM and CANDLE-Uno since its time and space complexity increases exponentially with respect to the number of parallel branches. \ours outperforms \pd and Piper at all but one GPU configuration. Moreover, the performance gap widens as the number of GPUs increases.


Our analysis reveals that we can attribute the widening performance gap to the pipeline depths greatly reduced by \ours compared to \pd and Piper for the multi-branch models. As we use more devices, the number of sequential pipeline stages tends to increase to achieve a higher throughput, particularly when the model size is too large to apply data parallelism at the cost of weight memory footprint and weight synchronization. With a larger number of stages, sequential pipeline schemes by generated by \pd or Piper suffer from extended warm-up and cool-down phases. Directly, these extended pipeline bubbles negatively affect training throughput. Indirectly, these bubbles increase activation memory footprints, which in turn impede effective model partitioning. We visualize this analysis in detail via a case study (see \S\ref{sec:case-study}).



\subsection{Search Time} \label{sec:search-time}

Table~\ref{tab:search-time} presents the search times by the three optimizers (\ours, \pd, and Piper) for the three models (Multi-Modal Transformer, DLRM, and CANDLE-Uno). The Multi-Modal Transformer-based model has two branches and the DLRM and CANDLE-Uno models have eight branches.

\ours is at least $9\times$ faster than the baselines irrespective of the models or GPU configurations. In addition, \ours's efficient partitioner produces a strategy within a minute for all configurations. The \spp baselines are much slower by comparison, and this search time discrepancy can be attributed in large part to the fact that the baselines rarely leverage DNN topology in expediting search. Note that Piper does not produce strategies
for the DLRM and CANDLE-Uno models for the aforementioned reasons. 

To see the large search space of each \spp baseline, it is helpful to approximate their time complexities. Let us consider a simple multi-branch model with each branch having $k > n$ operators, where $n$ is the number of branches. Recall that Piper considers model partitions in which cross-branch stages exist. This level of granularity of model partitions significantly increases the number of model partitions to examine. Piper's optimizer runs in $O(|\m{D}|^2)$ time (Appendix D in \cite{tarnawski2021piper}), where $\m{D}$ is the set of downsets (Definition 4.1 in \cite{tarnawski2021piper}). According to the definition, model partitions in which one stage spans multiple branches and all other stages are formed within a branch are valid candidates. Since we can choose one operator out of $k$ from each branch to form a cross-branch stage, the number of such model partitions is at least $|\m{D}| \geq \prod_{i=1}^{n} k = k^{n}$. Thus, Piper's time complexity is lower-bounded by $O(k^{2n})$. This time complexity implies that unless we employ a set of clever heuristics, Piper's time complexity can be significantly high for multi-branch DNNs.


On the other hand, \pd considers a converted DNN that linearizes all branches and the operators within. Thus, it deals with a single chain of operators, where the number of model partitions to consider is much smaller than Piper.


Still, \ours considers significantly fewer model partitions than \pd (and hence Piper) particularly when a given DNN features multiple branches. Instead of solving a single long chain of $nk$ operators as in \pd, \ours solves $n$ short chains of $k$ operators separately. As empirically shown in \Cref{fig:e2e-plot}, \ours barely demonstrates throughput degradation, which could have resulted from examining much fewer model partitions. Explicitly leveraging DNN topology in examining model partitions in search for a training strategy turns out to be critical to reducing the search space and time complexity.


\subsection{Different Numbers of Branches and \\Micro-Batch Sizes}

\begin{figure}
    \centering
    \includegraphics[width=0.48\columnwidth]{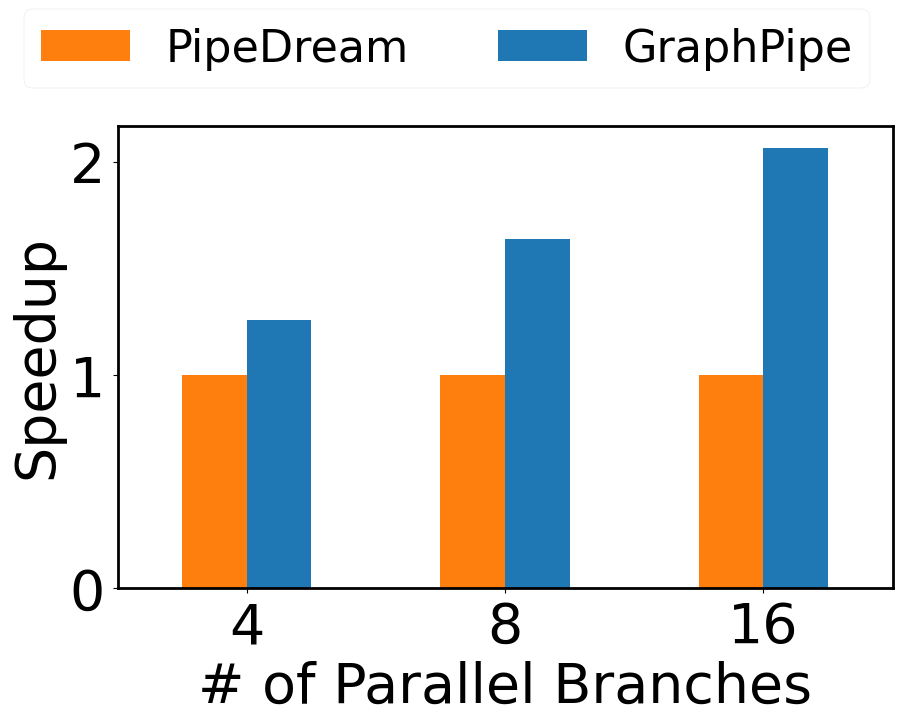}
    \includegraphics[width=0.48\columnwidth]{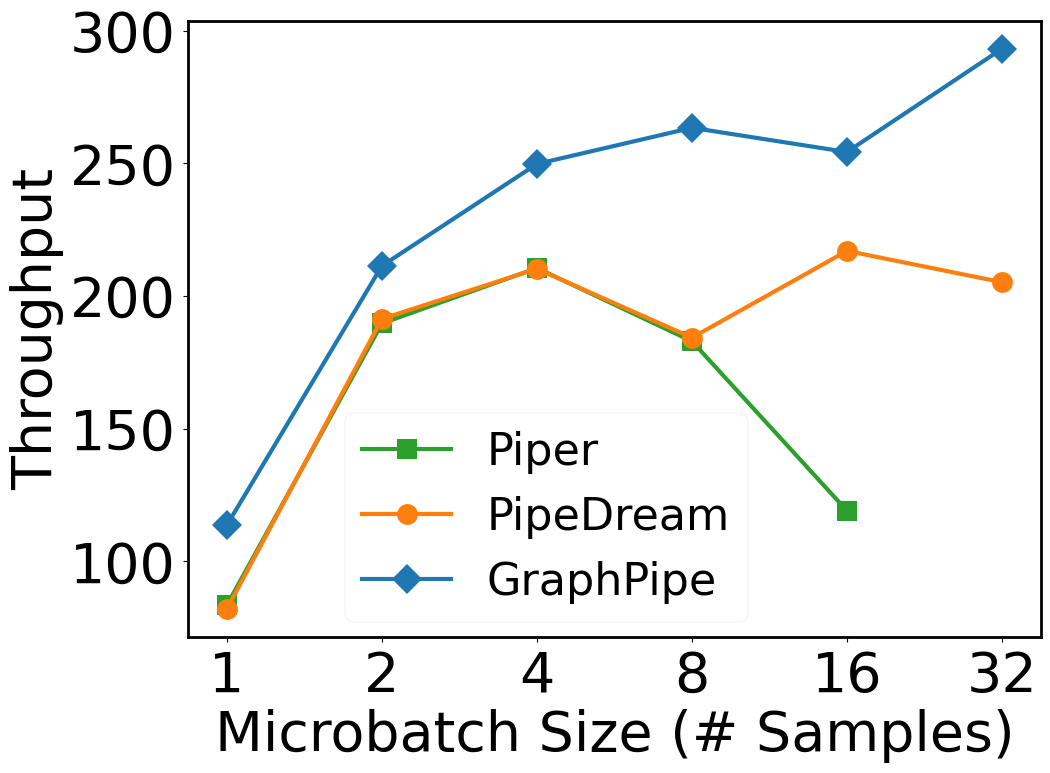}
    \caption{Throughput vs. different numbers of branches using 4, 8, 16 GPUs respectively (left). Throughput vs. different micro-batch sizes using 8 GPUs (right).}
    \label{fig:diff}
\end{figure}

Figure~\ref{fig:diff} shows the results of two experiments in which we change the number of parallel branches for the CANDLE-Uno model (left) and change the number of micro-batch sizes for the two-branch multi-modal Transformer-based model (right). The purpose of the experiments is to investigate the effects of main parameters on the performances of \ours and the \spp baselines (i.e., \pd and Piper).

The left sub-figure depicts the throughputs of different systems normalized by that of \pd with respect to the number of parallel branches for the CANDLE-Uno model.\footnote{Piper was not able to produce a strategy for the CANDLE-UNO model.}
We see that the performance gap achieved by \ours scales with the number of branches, reaching up to 2$\times$ at 16 branches. Intuitively, the performance gain mostly stems from the fact that \ours is able to reduce the pipeline depths at all configurations allowing concurrent execution of parallel branches, reducing the inefficient pipeline warm-up and cool-down phases significantly. The gain scales because the larger the number of branches, the larger the differentials of the phases between \ours and \spp. This experiment result demonstrates that (1) reducing pipeline depth is critical to training performance; and (2) \ours is better at it than \spp especially when multiple branches of non-negligible workload are present. The larger the number of branches in a given DNN to train, the more opportunities for \ours to exploit and reduce pipeline depths.

The right sub-figure depicts the throughput performances for the multi-modal Transformer-based model with four branches. We use a mini-batch size of 128 and eight GPUs. We intentionally fix a micro-batch size (instead of using the best ones chosen by the optimizers) in comparing the performances, for the purpose of examining the benefits (or harms) of using large micro-batch sizes. If increasing micro-batch size turns out to be beneficial, then it is worth reducing pipeline depth so as to reduce activation memory footprints, and in turn create room for using a larger micro-batch size.

\begin{figure*}[t!]
    \centering
    \includegraphics[width=0.89\textwidth]{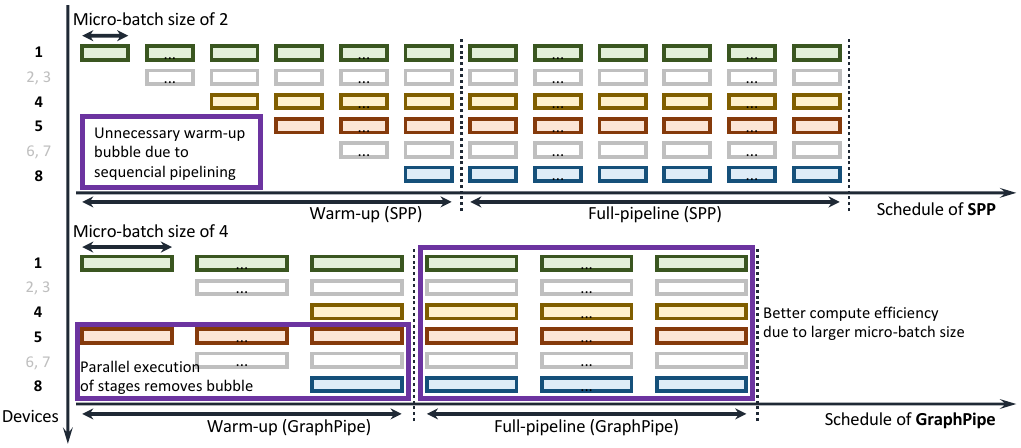}
    \caption{Pipeline schemes devised by \spp (top) and \ours (bottom). They produce an identical model partition. The selected micro-batch sizes are different: 2 (\spp) v.s. 4 (\ours), which results in a better compute efficiency for \ours. Both methods deem it unnecessary to employ data parallelism primarily because doing so would have split a smaller micro-batch size even further, which would have harmed compute efficiencies. The pipeline depths are also different: 8 (\spp) v.s. 4 (\ours), which results in a smaller pipeline depth for \ours. This improvement comes purely from the fact that \ours can produce a pipeline scheme that allows for concurrent execution of parallel branches.}
    \label{fig:case-study-schedule}
\end{figure*}

\begin{figure}[h!]
    \centering
    \includegraphics[width=0.7\columnwidth]{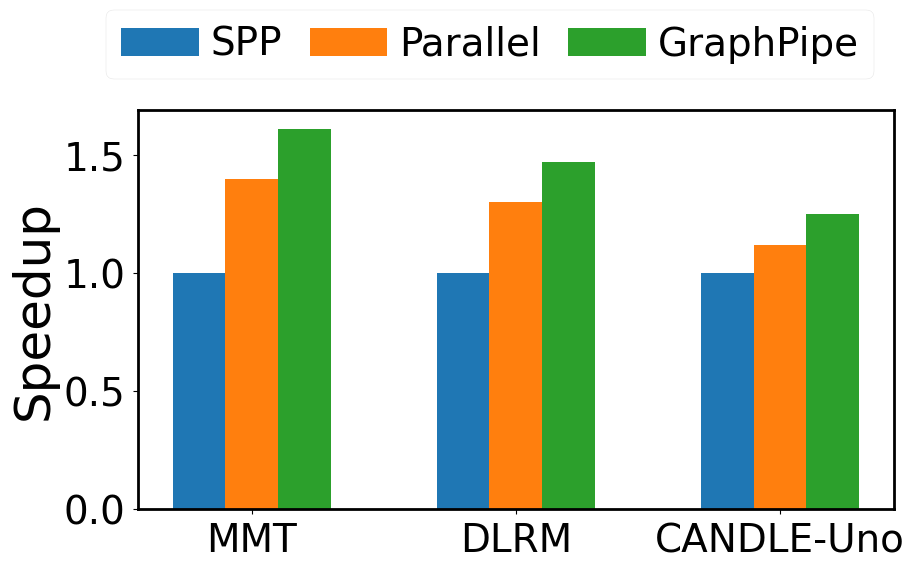}
    \caption{Ablation study: end-to-end throughput comparison of different strategies on different models.}
    \label{fig:ablation_study}
\end{figure}

We can observe the key role of reduced pipeline depth by \ours in improving throughput. For each micro-batch size, \ours always outperforms \spp. Since there is no difference in operational intensity with the same micro-batch size used for both \ours and \spp, the performance gap can be solely attributed to the difference in pipeline depth. The reduced pipeline depth by \ours leads to a shorter execution time for the warm-up and cool-down phases, hence a higher throughput.

\subsection{Ablation Study} 
\label{sec:ablation-study}

\Cref{fig:ablation_study} shows the breakdown of performance benefits of \ours from 1) parallel execution of stages and 2) increased micro-batch size from reduced memory footprint. In \Cref{fig:ablation_study}, "Parallel" is the strategy that allows parallel execution of stages with same micro-batch size with \spp while "GraphPipe" is the strategy that allows both parallel execution of stages and larger micro-batch size than \spp. Note that It is not possible to evaluate the strategy only with larger micro-batch size since the reduced pipeline depth from parallel stage execution enables larger micro-batch size. We evaluate throughput of each strategy with 32 GPUs. 

We observe that, compared to \spp, "Parallel" strategy achieves 1.12 - 1.40$\times$ speedup while "GraphPipe" strategy achieves 1.25 - 1.61$\times$ speedup. This result indicates that both performance benefits of \ours are crucial. We also find the consistent pattern of the strategy optimized by \ours across different models. In the following section, we explain how these two sources of performance gains contribute to the overall improvement achieved by \ours.

\subsection{Case Study} \label{sec:case-study}
\begin{figure}[t!]
    \centering
    \includegraphics[width=\columnwidth]{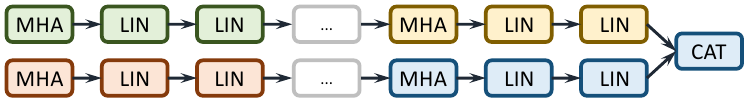}
    \caption{A synthetic Transformer-based two-branch DNN for case study. A sequence of one multi-head attention and two linear layers is repeated four times to compose a single branch. One concatenation layer combines two branches.}
    \label{fig:case-study-model}
\end{figure}

To clearly illustrate the advantages of \ours, we analyze the strategies it produces in comparison to \spp, using a simplified synthetic model for clarity.
We run both \ours and \spp optimizers, execute the strategies, and observe a 20\% throughput improvement by \ours over \spp. Our analysis finds that the aggregate gain comes from two sources, and the contributions are nearly equal. 

Figure~\ref{fig:case-study-model} depicts the two-branch Transformer-based model synthesized for the experiment. Each branch consists of four repeated sequences of one multi-head attention and two linear (dense) layers. The branches are merged by a concatenation operator.

Both \ours and \spp produce the identical model partition on a budget of eight devices. Each stage contains one multi-head attention and two linear layers. There are eight such stages, four per branch, except that one stage necessarily contains the concatenation operator.
A key difference between the two strategies, however, is the way the stages are pipelined. Figure~\ref{fig:case-study-schedule} depicts the pipeline schedules. Note that the pipeline depth for \spp is eight since all eight stages form a sequential pipeline. In contrast, the pipeline depth for \ours is four. The two branches are computationally-independent, hence stage $1+i$ and $5+i$ for $0 \leq i \leq 3$ can be executed in parallel, and this is precisely what the training strategy produced by \ours suggests. This concurrent execution reduces the warm-up phase by half in terms of number of micro-batches from eight to four. This warm-up phase reduction leads to 10\% performance improvement.

There is another subtle, yet key difference. Since \ours reduces the pipeline depth by half, the activation memory footprints for early stages are smaller for the \ours strategy. As a result, \ours can choose a micro-batch size from a wider range of candidates, and indeed selects a size of 4. The compute efficiency improvement from choosing a larger micro-batch size over \spp (which chooses a size of 2 due to larger activation memory footprints) leads to a larger number of samples processed per unit time. This means that when the pipeline operates at full capacity, it processes training samples at a faster rate for \ours than for \spp. Our measurements show that the gain from this compute efficiency improvement is 10\%. The two gain sources combined, \ours achieves 20\% higher throughput over \spp.

GraphPipe's performance improvement becomes more significant as 1) memory pressure and 2) the parallelism degree within a DNN increase. The selection of hardware influences memory pressure, subsequently affecting GraphPipe’s performance improvement, while it is common practice for the system to operate close to memory limits in DNN training. In contrast, the parallelism degree within a DNN is hardware independent, and therefore GraphPipe’s performance improvement over existing systems will maintain across different hardware platforms.
\section{Related Work}
\label{sec:related}


\textbf{Pipeline parallelism.} Existing DNN frameworks~\cite{TensorFlow, PyTorch, jia2019beyond, rajbhandari2020zero, shazeer2018mesh} employ sequential pipeline parallelism (\spp) where pipeline stages are strictly sequential. As we discuss in \Cref{sec:gpp}, \spp hinders parallel execution of computationally-independent components of a DNN and memory savings from reduced pipeline depth. While this limitation still exists as long as \spp is adopted, there are a variety of pipeline parallelism approaches to improve pipeline performance in other ways. These approaches fall into one of two paradigms: synchronous and asynchronous pipeline parallelism.

\emph{Synchronous pipeline parallelism} \cite{huang2019gpipe, narayanan2021memory, fan2021dapple, zheng2022alpa, unger2022unity} refers to a set of techniques in which the model parameters spread across devices are updated synchronously after every training iteration. The DNN training semantics is preserved, thus statistical convergence issues do not arise. But the synchronous updates fill and drain the pipeline periodically over iterations, hurting throughput. Our graph pipeline parallelism mitigates this issue by reducing pipeline bubbles better than sequential pipeline parallelism.

\emph{Asynchronous pipeline parallelism} \cite{narayanan2019pipedream, narayanan2021memory, tarnawski2021piper,wpipe} refers to a set of techniques in which the model parameters spread across devices are updated asynchronously. Although this mode may suffer from statistical convergence issues as devices execute their stages using out-of-sync model parameters, it keeps the pipeline full at nearly all times. Graph pipeline parallelism helps us reduce total device memory usage, thus use a larger micro-batch size to execute operators at a higher operational intensity compared to sequential pipeline parallelism. This enables us to process training data faster while the pipeline is full.

\textbf{Multiple pipeline stages per device.} In the pipeline parallel techniques above, each device contains only one pipeline stage. It has been shown that assigning multiple non-contiguous stages to a device can reduce pipeline bubbles \cite{narayanan2021efficient, lamy2022breadth} and reduces memory consumption imbalances across stages~\cite{li2021chimera, hanayo}. Earlier work GEMS~\cite{gems} has a similar idea but does not utilize the pipeline well --- devices are idle for most of the time and waiting for results from other stages. These techniques are orthogonal to graph pipeline parallelism, and thus can be applicable upon some modifications.

\textbf{Data parallelism.} Data parallelism \cite{valiant1990bridging, krizhevsky2014one, li2014scaling, goyal2017accurate, mudigere2021high} is one of parallel DNN training techniques in which every device has a local copy of a DNN to train and a batch of training data is split across devices. Each device updates its model parameters based on its share of training data and synchronizes the parameters periodically with other devices. In our work, we apply data parallelism within a pipeline stage to which we assign multiple devices, in order to balance stage execution times in a more fine-grained manner compared to applying pipeline parallelism only.

\textbf{Automatic DNN parallelism.} There are a number of automated approaches~\cite{zheng2022alpa, unger2022unity, jia2019beyond, tarnawski2021piper, narayanan2019pipedream, mirhoseini2017device, wang2019supporting} combining data, pipeline, and tensor parallelisms~\cite{shoeybi2019megatron}. Existing works first partition a DNN into sequential pipeline stages (\spp) and then apply data and tensor parallelism to each stage. \ours follows this same high-level process as well. However, the key difference is that it generalizes stage partitioning to produce graphical stages and exploit concurrent execution opportunities from DNN structures (i.e., parallel branches). Note that it is also feasible to combine our approach with tensor parallelism by adding a subroutine of applying tensor parallelism (e.g., intra-op pass in Alpa~\cite{zheng2022alpa}) in our partitioner while our scheduler and runtime are already compatible.

\iffinal
\section{Conclusion}

We have developed \emph{graph pipeline parallelism} where pipeline stages form a directed acyclic graph whose edges indicate execution orders of forward and backward passes in pipeline-parallel DNN training. This design encourages \emph{concurrent} execution of parallel branches for superior performance. We have also developed a distributed system \ours, and through experiments using three multi-branch models, showed that \ours achieves up to 1.61$\times$ higher training throughputs and $>9\times$ faster solution search times over existing baselines that operate in a strictly sequential manner.

\section*{Acknowledgement}
We would like to thank members of Catalyst group at CMU for their helpful comments on our work and manuscript. We would also like to thank the anonymous ASPLOS reviewers for constructive feedbacks. This work was partially supported by the National Science Foundation under grant numbers CNS-2147909, CNS-2211882, and CNS-2239351, along with gift awards from Amazon, Cisco, Google, Meta, Oracle, Qualcomm, and Samsung. Additional support was provided by the Real Time Machine Learning (RTML) DARPA project.

\fi 


\appendix
\section{Appendix}
\subsection{Generalized Per-stage $k$F$k$B Schedule}
\label{sec:nfnbrule}
The $k_x$F$k_x$B schedule of stage $S_x$ is determined by
\begin{align*}
    \mathrm{argmin}_{k_x} \max_{(S_x, S_y) \in \m{V}_S} \mathrm{ComputeInFlight}(k_x, b_x, k_y, b_y, i_y),
\end{align*}
where $i_y, b_y$ are the number of in-flight samples and micro-batch size for stage $S_y$. $\mathrm{ComputeInFlight}(k_x, b_x, k_y, b_y, i_y)$ is computed according to Table~\ref{tab:nfnb_rule}:

\begin{table}[h]
    \centering
    \begin{tabular}{c|c}
        \toprule
        Condition & Result \\
         \midrule
        $\max\{b_x, b_y\} < k_xb_x < k_yb_y$ & $ i_y + 2\max\{b_x, b_y\}$ \\
        $\max\{b_x, b_y\} = k_xb_x < k_yb_y$ & $i_y + \max\{b_x, b_y\}$ \\
        $b_x \le b_y < k_yb_y < k_xb_x$ & $i_y+k_xb_x-k_yb_y+2b_y$ \\
        $b_x \le b_y = k_yb_y < k_xb_x$ & $i_y+k_xb_x$ \\
        $b_y \le b_x < k_yb_y < k_xb_x$ & $i_y+k_xb_x-k_yb_y+2b_x$ \\
        $b_y \le b_x = k_yb_y < k_xb_x$ & $i_y+k_xb_x$ \\
        $\max\{b_x, b_y\}=k_yb_y=k_xb_x$ & $i_y+k_yb_y$ \\
        $\max\{b_x, b_y\}<k_yb_y=k_xb_x$ & $i_y+2\max\{b_x, b_y\}$ \\
        $b_x \le k_xb_x < b_y \le k_yb_y$ & $i_y+b_y$ \\
        $b_y \le k_yb_y < b_x \le k_xb_x$ & $i_y+k_xb_x-k_yb_y+b_x$ \\
        \bottomrule
    \end{tabular}
    \caption{Computation of the number of in-flight samples.}
    \label{tab:nfnb_rule}
\end{table}

\subsection{DNN Model Configurations}
\label{sec:model_config}

The Multi-Modal Transformer-based model (MMT) for which we evaluate \ours consists of four parallel branches concatenated at the end and each branch consists of eight Transformer layers (32 layers in total). Here, the input sequence length is 256. Each transformer layer has a hidden size of 1024, an embedding size of 1024, and 16 attention heads. The hidden size for a feed-forward layer following the attention layer has a hidden size of 4096.

The DLRM model for which we evaluate \ours consists of seven branches for dense features and seven branches for sparse features (embedding layers); these branches are concatenated at the end. Each branch for dense features includes four feed-forward layers. The hidden size of dense features and the following feed-forward layers is 4096. For sparse features, its hidden size is 64 and the embedding bag size is 100; embeddings in a single bag is concatenated. The number of entries in an embedding table is 1 million. Feed-forward layers post-processing the interaction also have the hidden size of 4096.

The CANDLE-Uno model for which we evaluate \ours consists of seven branches, each of which includes four feed-forward layers. All feed-forward layers have a hidden size of 4096.

For our end-to-end evaluations, we use the following ranges of mini-batch sizes for each device count such that the system operates close to the memory limit:
\begin{table}[h]
    \centering
    \begin{tabular}{c|ccc} \toprule
       \# Devices & MMT & DLRM & CANDLE-Uno \\ \midrule
        4 & 64 & 256 & 4096 \\
        8 & 128 & 512 & 8192 \\
        16 & 256 & 1024 & 16384 \\
        32 & 512 & 2048 & 32768 \\
        \bottomrule
    \end{tabular}
\end{table}

We sweep over all possible micro-batch sizes given mini-batch sizes for each model to maximize training throughput.

\subsection{Sequential DNN Evaluation}
\label{sec:seq_dnn_eval}

We evaluate \ours with baselines and confirm that \ours match performance of baselines when the workload is sequential DNN. We measure throughput on the sequential Transformer model with the same configuration with the MMT model above.

\begin{table}[h]
    \centering
    \begin{tabular}{c|ccc} \toprule
       \# Devices & Piper & PipeDream & Ours \\ \midrule
        4 & 113.94 & 113.2 & 113.6 \\
        8 & 187.92 & 190.1 & 189.5 \\
        16 & 286.83 & 277.00 & 287.52 \\
        32 & 609.76 & 592.4 & 608.82 \\
        \bottomrule
    \end{tabular}
\end{table}

\bibliographystyle{plain}
\bibliography{references}

\end{document}